\documentclass[a4paper,10pt,twoside]{cpc-hepnp}
\usepackage{multirow}
\usepackage{multicol}
\usepackage{graphicx}
\usepackage{booktabs}
\usepackage{amssymb,bm,mathrsfs,bbm,amscd}
\usepackage[tbtags]{amsmath}
\usepackage{lastpage}
\usepackage{hangcaption}

\renewcommand{\citet}[1]{\textsuperscript{\cite{#1}}}

\begin{document}

\fancyhead[co]{\footnotesize Sun Xiao-dong~ et al: The Online Data
  Quality Monitoring System at BESIII}

\footnotetext[0]{Received  2011}

\title{The Online Data Quality Monitoring System at
BESIII\thanks{Supported by the Knowledge Innovation Project of CAS
under Contract Nos. KJCX2-YW-N29 }}

\author{%
      Sun Xiao-Dong$^{1,2;1)}$\email{sunxd@mail.ihep.ac.cn}%
\quad Hu Ji-Feng$^{1,2;2)}$\email{hujf@mail.ihep.ac.cn}%
\quad Zhao Hai-Sheng$^{1,2)}$\\
\quad Ji Xiao-Bin$^{1}$
\quad WANG Yi-Fang$^{1}$
\quad ZHENG Yang-Heng$^{2}$\\
\quad LIU Bei-Jiang$^{1}$
}
\maketitle

\address{%
  $^1$ Institute of High Energy Physics, Chinese Academy of Sciences, Beijing 100049, China\\
  $^2$ Graduate University of Chinese Academy of Sciences, Beijing 100049, China\\
}

\begin{abstract}
  The online Data Quality Monitoring (DQM) plays an important role in
  the data taking process of HEP experiments. BESIII DQM samples data
  from online data flow, reconstructs them with offline reconstruction
  software, and automatically analyzes the reconstructed data with
  user-defined algorithms. The DQM software is a scalable distributed
  system. The monitored results are gathered and displayed in various
  formats, which provides the shifter with current run information
  that can be used to find problems early. This paper gives an
  overview of DQM system at BESIII.
\end{abstract}



\begin{multicols}{2}

\section{Introduction}
BESIII is a detector operating on Beijing electron-positron collider
(BEPCII) at the Institute of High Energy Physics (IHEP) of the Chinese
Academy of Sciences in Beijing. With design luminosity of ~$10^{33}
cm^{-2}s^{-1}$~of BEPCII, BESIII will collect large data samples so
that $\tau$-charm physics can be studied with high
precision\citet{bes3}.

The peak luminosity has exceeded $6\times10^{32}cm^{2}s^{-1}$
recently. At so high luminosity, it's essential to monitor the status
of the BESIII hardware and to determine the quality of acquired data
in time. In the online environment, the DAQ (Data AcQuisition) system
uses all the acquired data to get preliminary information of
sub-detectors. But because of the multi-beam bunches and the pipelined
readout electronics system, some information such as event start time
cannot be obtained in the DAQ system. Offline reconstruction can give
much more information about the detector performance and data quality,
but its results will not be available until several days after data is
taken. To monitor data quality both in an accurate way and in real
time, the online Data Quality Monitoring system (DQM) is
developed. DQM fully reconstructs part of the acquired data, which is
sampled randomly from online data flow, using the offline full
reconstruction software. The monitored results on detector status
and data quality will be available only a few minutes after one run
begins.

\section{Properties and operating environment}

The BESIII online DQM system consists of 8 nodes: 5 IBM eServerBlade
HS20, each with dual 3.0 GHz Xeon CPUs, and 3 PCs.  All nodes have SLC
4.6 operating system installed. 5 HS20 nodes process the event data,
including event reconstruction and analysis. one PC is used as DQM
system server to provide system management and DQM services, such as
histogram merge and histogram display; one PC deals with Event
Display; the other one is the backup machine.

The communication between DQM and DAQ is only handled by high speed
network connection. Events copied from the online data stream are
transfered one by one to DQM machines through the network and does not
influence the online data flow. Only a part of the events in data are
sampled. The sampling rate depends on the processing capacity of the
DQM programs. Most of the time consuming process is the data
reconstruction. As for now, the maximum sampling rate can reach ~360
Hz. And the ratio of signal events (Bhabha, Dimuon and Hadron events)
is about 40$\sim$60 Hz.

The BESIII DQM software framework was developed based on the framework
of the ATLAS DQMF\citet{atlas} and BESIII offline software system
(BOSS). The work flow is redesigned to ensure the separation from the
DAQ software system and automatic control under BESIII
enviorment. Only a data server is required to run in the online DAQ
system. The data server samples data from the online data flow and
then sends the data to DQM processes by TCP connections. The design
ensures the safety of the DAQ system. And DQM system integrates with
BOSS tightly. Almost all the offline algorithms can run under the DQM
environment with minor or no modification. DQM system is a distributed
system. There are totally 20 DQM main processes running on the HS20
nodes. The histograms generated by each main process are merged by
Histogram Merger process in real time. Then the merged histograms can
be checked during the data taking and are stored in root files after a
run finished. The detector properties such as TOF time resolution, MDC
momentum resolution obtained from each run are stored in the DQM
database for stability check of the detector.

\section{The architecture of DQM}
\begin{center}
\includegraphics[width=8.cm, height=6cm]{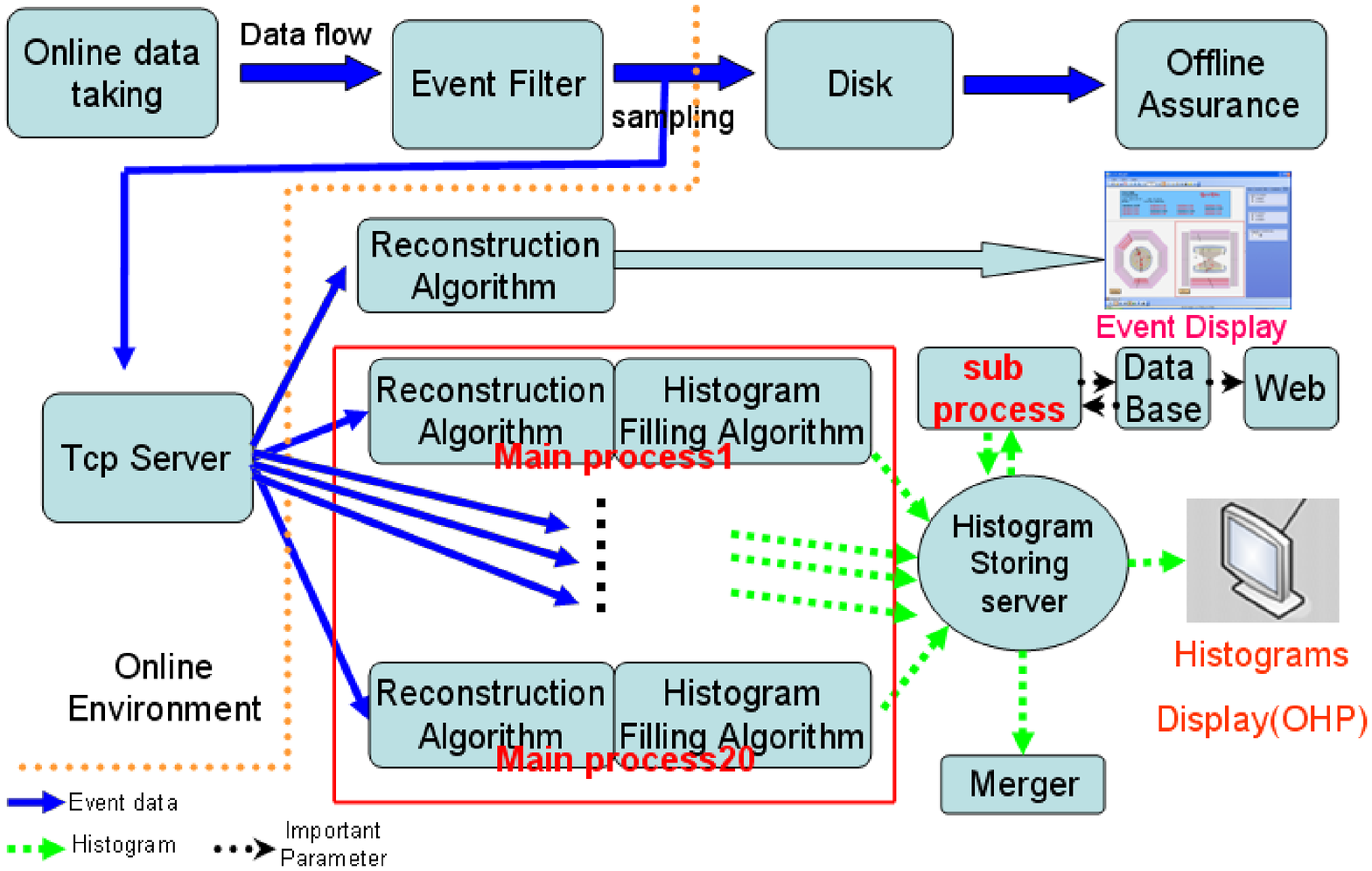}
\figcaption{\label{fig1}   The framework of DQM (color online) }
\end{center}
The framework of DQM\cite{hujf} is shown in Fig.~\ref{fig1}. DQM
system consists of 6 main parts, i.e. DQM Server, DQM clients, Merger,
Histogram Storing server, DQM database and information display
programs. DQM Server is used to get data from online data flow. DQM
clients invoke reconstruction algorithms and data analysis algorithms
to reconstruct events and fill histograms. Merger is used to merge
histograms from different DQM clients. Histograms Storing server is
where all histograms are stored. DQM database stores parameters that
reflect detector properties and data quality of each run. Information
display programs are used to display information that is useful for
monitoring data quality.

\subsection{Basic work flow of DQM}

The event fragments received from different sub-detectors are
assembled into full events and filtered (Event Filter\citet{ef}). A
TCP server, which is called DQM Server, is used to sample (copy)
events passed the Event Filter and then deliver them to different DQM
clients.

DQM clients reconstruct events to get basic information of each
sub-detector. A small part of the reconstructed events are displayed
directly by Event Display program, while the other are used to fill
first-class histograms (histograms filled with information in event
level). The first-class histograms from different DQM main processes
are merged by a Histogram Merger after they are published to Histogram
Storing server. Some of the merged histograms will be displayed by
Online Histogram Presentation program (OHP) directly and some will be
used to generate second-class histograms (histograms filled with
information extracted from first-class histograms) to be displayed by
OHP. Some global parameters of sub-detectors, such as momentum
resolution, time resolution, event vertex and so on, are extracted
from the histograms in the end of the run and stored into DQM
database.

\subsection{DQM Clients/DQM main processes }

There are 22 DQM clients currently, including main process,
sub-process and event display. 20 DQM clients are DQM main processes
which fully reconstruct the events and then fill first-class
histograms using the information in the reconstructed events. DQM
sub-process used to handle first-class histograms in the end of a
run. Event display program fully reconstructs the events and displays
it visually in real time, as shown in Fig.~\ref{fig2}.

Framework of DQM main processes is shown in Fig.~\ref{fig3}. A DQM
main process can be divided into 5 functional units. The TCP client is
used to fetch events from DQM Server. Then the interface program
unpacks the data and supplies events in the format suitable for
reconstruction algorithms. DQM uses the offline reconstruction
algorithms to fully reconstruct the events. Histogram-filling
algorithms (user defined algorithms) use the information of the
reconstructed events to fill histograms. Histogram publish program
publishes histograms to Histogram Storing server.
\begin{center}
\includegraphics[width=7.8cm, height=6cm]{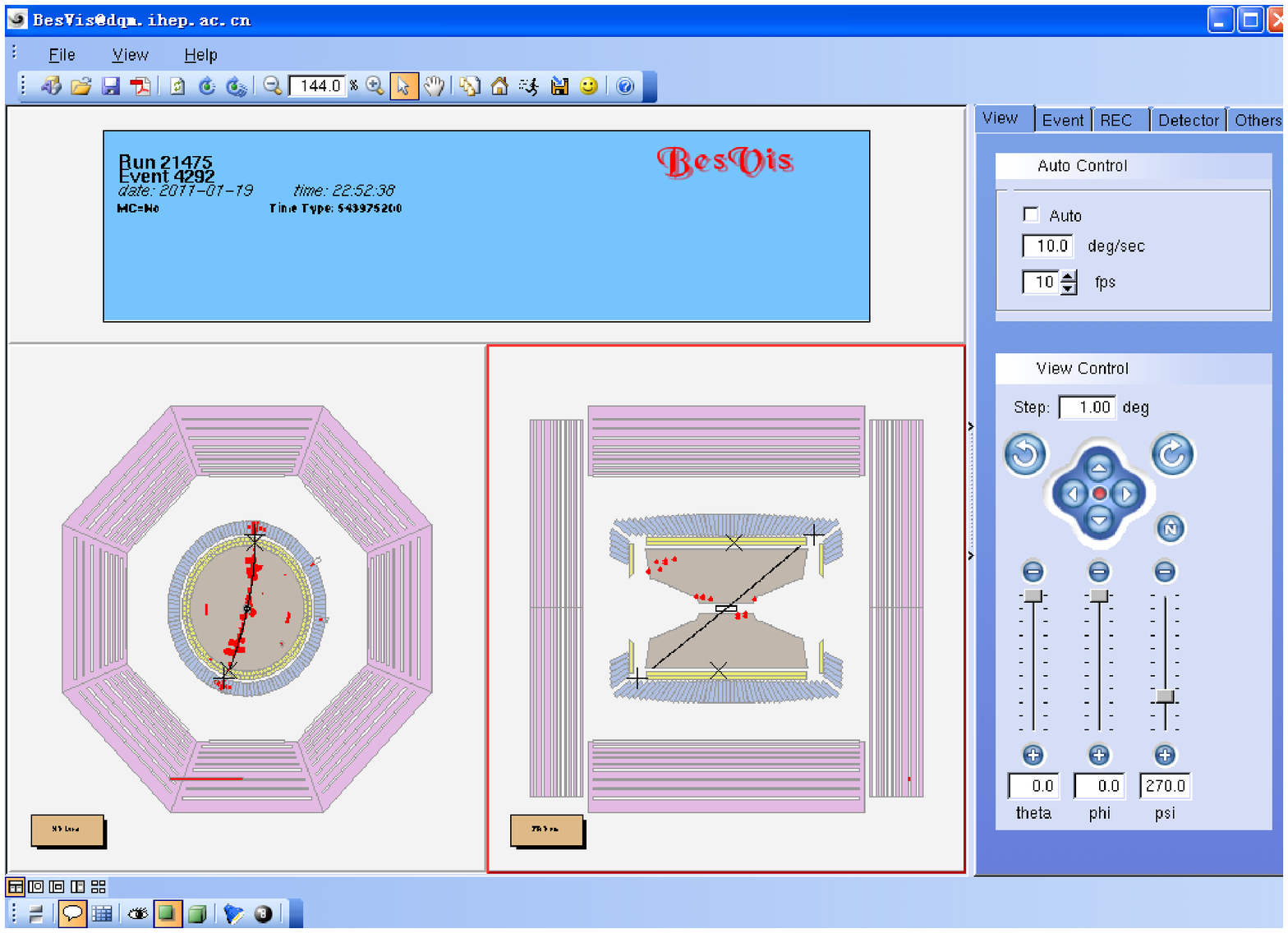}
\figcaption{\label{fig2} A Bhabha event displayed by Event Display
  (color online) }
\end{center}

\begin{center}
\includegraphics[width=7.8cm, height=4.8cm]{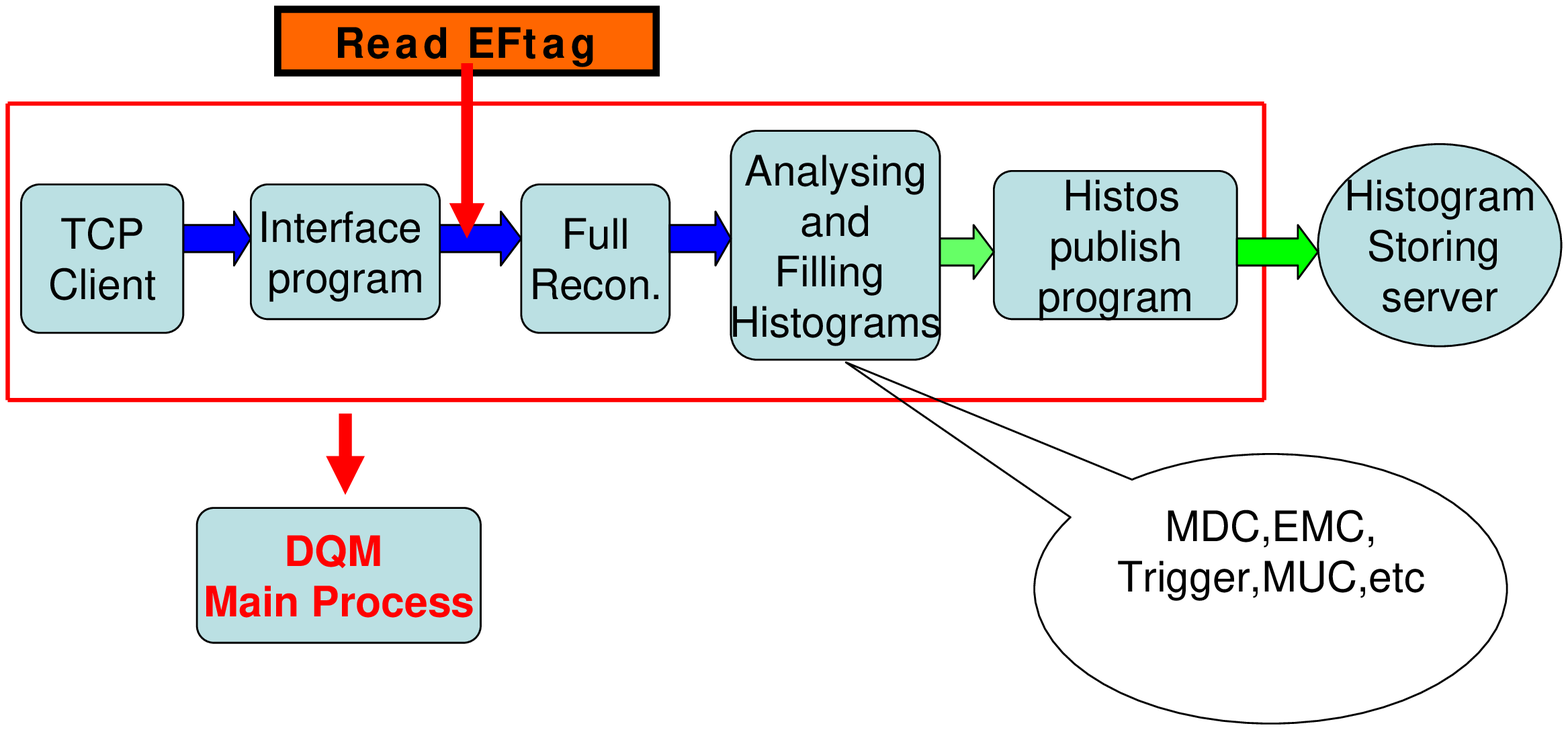}
\figcaption{\label{fig3}   Framework of DQM main processes (color online) }
\end{center}

To improve the processing ability, the KalFitAlg
algorithm\citet{kalfit} used for track fitting is not used in DQM
online reconstruction, which is different from offline
reconstruction. As a result, the time spent on event recontruction
decreases almost 50\% while the precision of DQM result is slightly
worse than offline result. Another important difference between DQM
reconstruction and offline reconstruction is that DQM reconstruction
uses the calibration constants from recent runs.

In order to further improve the process capacity of DQM system,
event tags given by Event Filter are checked by a specific algorithm,
which is executed before the event reconstruction and only events with
specific Event Filter tag are reconstructed. The processing rate of
DQM improved 44\% under current configuration.

\subsection{Histogram-filling algorithms (user defined algorithms)}
DQM main processes invoke user-defined histogram-filling algorithms to
fill histograms. The algorithms are flexible and independent of each
other. Each of the histogram-filling algorithms can be added or
removed easily, which ensures the extensibility of DQM system.

Many histogram-filling algorithms have been developed to fulfill the
task of monitoring.  A special event tag algorithm is used for event
classification using the reconstructed information. Each event
processed by DQM is tagged as Bhabha, Dimuon, Hadron, Cosmic Ray and
so on for later use.

Main histogram-filling algorithms are used to monitor the sub-system
of BESIII, including MDC(Main Drift Chamber), TOF(Time-of-Flight
counters), EMC(Electro-Magnetic Calorimeter), MUC(Muon Counter), and
trigger system. All these algorithms can use the event tags from event
tag algorithm to fill histograms with some kind of events
only. Furthermore, several histogram-filling algorithms related to
some special physics channels, such as inclusive $K_S$, inclusive
J/$\psi$ in $psi^\prime$ data, D meson in $\psi^{\prime\prime}$ data,
and so on, are developed, which give a more physics to monitor the
data.

Each of the detector-related alogithms has a sub-algorithm to fill
second-class histograms using the first-class histograms it
filled. DQM sub-process invokes sub-algorithms to extract information
in the merged first-class histograms and get parameters such as
detecting efficiency, noise ratio, energy resolution, etc. and then
fills second-class histograms with these parameters. Some important
parameters are stored into DQM database which is a subset of BESIII
Offline Database\cite{database}. These parameters are fetched back to
draw the historical curve in order to monitor the run stability.

\begin{center}
\includegraphics[width=7.8cm, height=4cm]{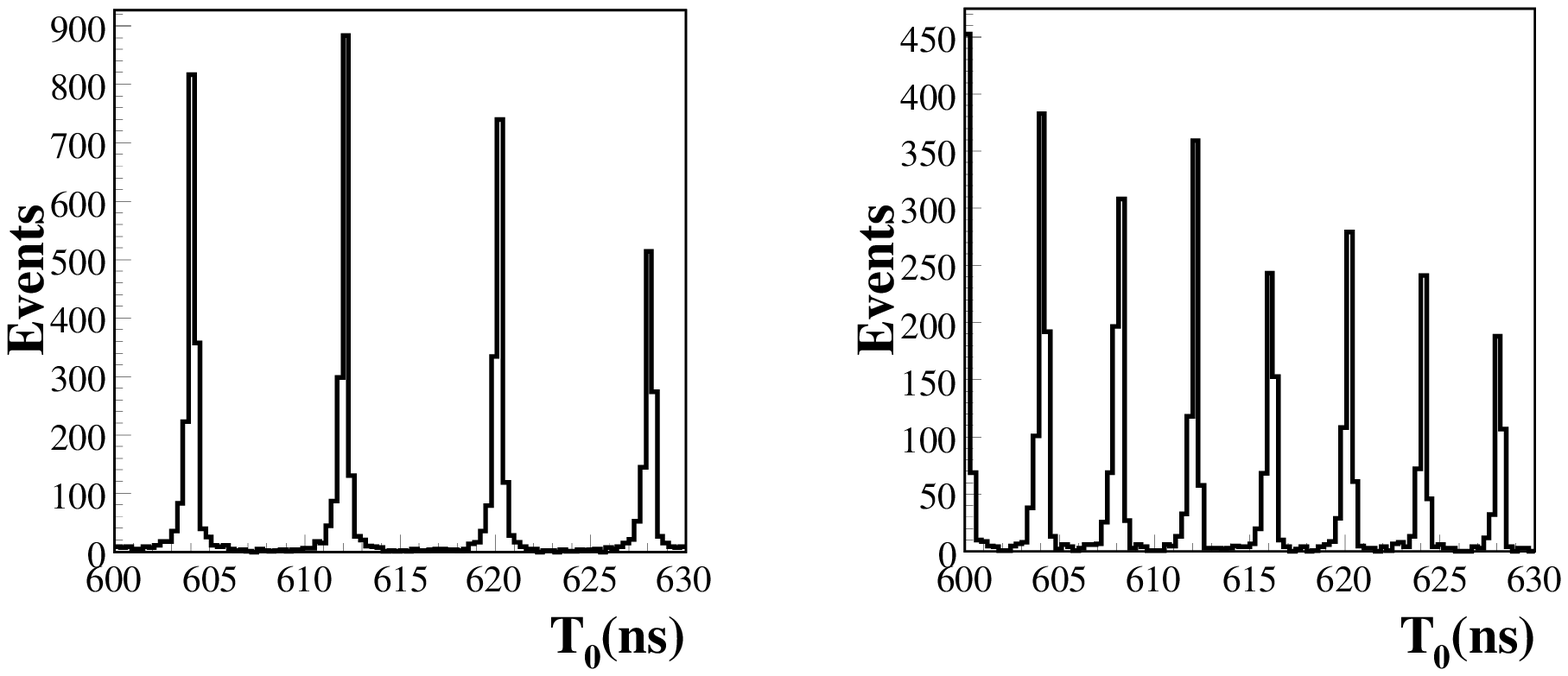}
\figcaption{\label{est}   Event start time }
\end{center}

Integral luminosity, event start time, event vertex, etc are also
monitored in the DQM system, which are sensitive to the beam
condition. The accelerator monitoring system can get them from
DIM\cite{dim}(Distributed Information Management System) either.
Fig.~\ref{est} is an example of event start time with different beam
intervals (8ns left, 4ns right). The shift can found the change of the
beam interval quickly during data taking.

\section{Cooperation of DQM processes}
DQM system is a completely automatic system in the data taking.  The
flow control system, based on TCP connections, controls different
components of the DQM system to ensure all of them taks right action
at right time.
	
When taking data during a Run, DQM Server will send events to DQM main
processes by TCP connections automatically. Once receiving events, DQM
main processes will reconstruct them and generate first-class
histograms.
	
At the end of a run, DQM main processes will publish all histograms of
this Run into Histogram Storing Server and then send a TCP signal to
Merger. After receiving the TCP signals from all 20 DQM main
processes, Merger will merge the histograms in Histogram Storing
Server, and then send a TCP signal to DQM sub-process. After that, DQM
sub-process will fetch the first-class histogram from Histogram
Storing Server, generate second-class histogram, publish second-class
histogram to Histogram Storing Server and then send a TCP signal to
DQM Histogram Storing Server. At last DQM Histogram Storing Server
will store all histograms of the Run into a ROOT file after receiving
the TCP signal from DQM sub-process.
	
All TCP connections are protected by time-out mechanism. The DQM
processes with TCP connections will wait for TCP signals before taking
further step. But if the TCP signals can not be received after a fixed
time, they will take action without the TCP signals so that when some
DQM processes are broken, the other processes can work still.

Several daemon processes are running to recover DQM system from
unexpected errors.  Once a DQM process crashes, it will be restarted
automatically.

\section{Information display}

DQM results are viewed in three ways: Event Display, Histogram
Display, and Web Display.

The Event Display program is modified according to the offline version
in BOSS. The reconstructed events can be displayed automatically in
real time. Fig. ~\ref{fig2} shows a Bhabha event displayed by Event
Display program. Two tracks can be seen clearly.

Histograms from different user-defined algorithms can be viewed easily
using tools OHP or OHD. Important histograms of each sub-detectors and
typical physics processes are displayed on OHP.  All the physical
variables filled into these histograms are shown in table 1.

%

\begin{center}
\begin{tabular}{p{4em}p{16.5em}}
  \hline
  separated parts & histograms (most given X axis only, Y is number of Bhabha events by default) \\
  \hline
  MDC & Momentum of $e^+$ and $e^-$ ,Residual\citet{mdc} distribution, event start time, dE/dx, $\phi$ and $cos\theta$ of $e^+$ and $e^-$ ,  drift time in inner/outer chamber \\
  TOF & ¦ $\Delta$T of Barrel/Endcap, East Barrel/West Barrel/Endcap hit map, barrel z of the hit position, time resolution of Endcap, time difference between upper  and lower TOF \\
  EMC & Shower energy deposited in Barrel/Endcap EMC, Shower $\phi$ in Barrel/Endcap, Shower ID($\theta$) \\
  MUC & ¦ $\phi$ vs. $\cos\theta$ for all events, fired layers in MUC of tracks, event No. vs. number of MUC hits, acollinear angle distribution of the momentum of dimu tracks \\
  Trigger & Fired ADC number for Barrel/Endcap, scintillator ID for Barrel/Endcap, long track hit map, trigger channel, trigger condition \\
  Physics & $\phi$ of $e^+$ and $e^-$  for Barrel, $cos\theta$ of $e^+$ and $e^-$ \\
  \hline
\end{tabular}
\
\tabcaption{ \label{tab1}  Histograms displayed on OHP to be checked by shift personnel.}
\end{center}

\begin{center}
\includegraphics[width=7.8cm, height=5.8cm]{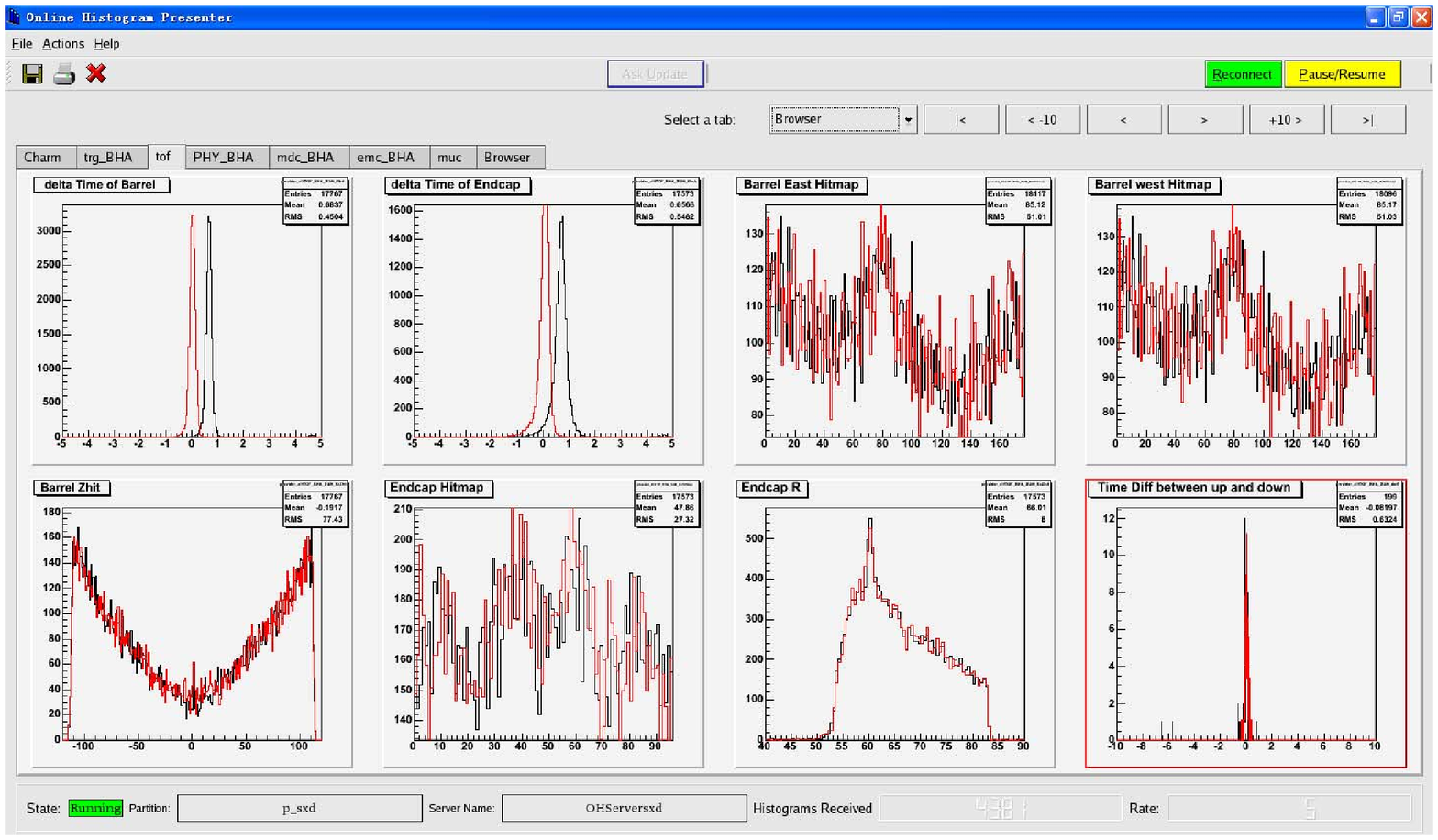}
\figcaption{\label{fig4}   An example of displaying histograms by OHP (color online) }
\end{center}

Most of the histograms shown on OHP are displayed with reference
histograms, as shown in Fig.~\ref{fig4}. The reference histograms come
from a recent good run. The shifter can find problem easily from OHP
when it occur. For example, first two histograms in Fig.~\ref{fig4} do
not agree with their references. Careful check points out that it is
caused by the accelerator problem.

OHD is a tool which can be used to check all the histograms of current
run by experts on sub detectors and other people who are
authorized. ROOT can also be used to check the historical histograms
stored into ROOT files.

Important parameters of detectors in each Run stored in DQM database
can be checked in tables or by histograms on the web.  People can
check the momentum resolution and space resolution of MDC, time
resolution of TOF, energy resolution of EMC, peak value of shower
energy deposited in EMC for $e^+$ and $e^-$ and the mean value of
x,y,z of the event vertex. The integral luminosity of each run are
calculated with Bhabha events and Di-photon events separately, and can
also be checked on the web.

\section{Summary}

DQM system has been developed and implemented in BESIII. After tuning
and updating carefully for the first several months, DQM has been
running continuously and steadily during BESIII data taking process
for more than two years since July, 2008. DQM system can monitor the
sub-detector and trigger system in much more physical and accurate way
than the other online monitoring system. DQM has become an essential
part of the whole BESIII data quality monitoring system. Together with
the online DAQ monitoring system, DQM ensures the successful and
robust data taking and physics analysis on BESIII.

\acknowledgments{The authors would like to thank Zhu Yong-Sheng and He
  Kang-Lin for helpful discussions and suggestions, and acknowledge
  Tian Hao-Lai, Zou Jia-Heng, Wu Ling-Hui, Sun Sheng-Sen, Liu
  Chun-Xiu, Xie Yu-Guang, Cao Guo-Fu for their contributions on part
  of DQM algorithms.}

\end{multicols}

\vspace{10mm}
\vspace{-1mm}
\centerline{\rule{80mm}{0.1pt}}
\vspace{2mm}

\begin{multicols}{2}

\end{multicols}

\clearpage

\end{document}